\documentclass[aps,jcp,preprint,superscriptaddress,letterpaper]{revtex4}

\usepackage[utf8]{inputenc}

\usepackage{amsmath, amssymb}
\usepackage{tabularx,booktabs,array,dcolumn}
\usepackage{setspace}
\usepackage{vmargin}
\usepackage{graphicx,psfrag,subfigure}
\usepackage{xcolor}

\usepackage{marvosym}

\usepackage[all]{xy}

\usepackage{tikz}
\usetikzlibrary{calc}
\usetikzlibrary{patterns}

\usepackage{threeparttable}
\usepackage{rotating}
\usepackage{multirow}

\newcommand{\dd}{\mathrm{d}}

\def\nuc{\text{nuc}}
\def\el{\text{el}}
\def\hrho{\hat{\rho}}
\def\dd{\text{d}}

\def\rot{{[\text{rot}]}}

\begin{document}

\title{
Orientational decoherence within molecules and emergence of the molecular shape
}

\author{Edit M\'atyus}
\email{matyus@chem.elte.hu}
\affiliation{Institute of Chemistry, E\"otv\"os Lor\'and University, P\'azm\'any P\'eter s\'et\'any 1/A, 1117 Budapest, Hungary}

\author{Patrick Cassam-Chena\"i}
\email{cassam@unice.fr}
\affiliation{Universit\'e C\^ote d'Azur,  CNRS, LJAD, UMR 7351, 06100 Nice, France}

\date{\today}

\vspace*{1.5cm}
\begin{abstract}
\noindent %
The question of classicality is addressed in relation with 
the shape of the nuclear skeleton of molecular systems.
As the most natural environment, the electrons of the molecule are considered as continuously monitoring agents for the nuclei. For this picture, an elementary formalism of decoherence theory is developed and numerical results are presented for few-particle systems. 
The numerical examples suggest that the electron-nucleus Coulomb interaction is sufficient for inducing a blurred shape with strong quantum coherences in compounds of the lightest elements, H$_2$, D$_2$, T$_2$, and HeH$^+$. 
\end{abstract}

\maketitle 

\section{Introduction}
\noindent%
We would like to better understand the status of the permanent chemical observables
by starting from a fully quantum mechanical description of a molecule, \emph{i.e.,} by including both the electrons and the atomic nuclei in the quantum treatment.
The present work is concerned with the recognition of elements of the 
classical molecular shape. A fundamental difficulty is associated with the fact that the classical shape (unless spherically symmetric) breaks the rotational symmetry of the molecular wave function of an isolated molecule \cite{Ma19review}.

There exist quick shortcuts to this problem. First (case~I), pieces of information regarding the molecular shape 
(and structure) can be obtained by \emph{fitting}
an effective model Hamiltonian to the isolated molecule's rotational spectrum \cite{molstruct2010}. 
Second (case~II), in many chemical experiments the molecules are in some 
environment, and thus, the isolated-system symmetries are not longer relevant.

Nevertheless, we remain interested in the original theoretical problem, that had became famous as the molecular structure conundrum \cite{Wo76,WoSu77,ClDi80,Primas81,We84,Lo89,SuWo05}: 
we consider an isolated molecule and aim to explore
the furthermost point regarding the molecular shape (and structure) 
without considering any kind of an environment (as in case~II)
or making any \emph{a priori} assumption on the hierarchical separation in the spatio-temporal behaviour of the internal dynamics to build model Hamiltonians (as in case~I). 

But we have said that the classical molecular shape breaks a fundamental symmetry of 
the isolated quantum system. So, how can we expect to see the emergence of a symmetry breaking feature without actually breaking the symmetry?

\begin{figure}
\includegraphics[scale=1.]{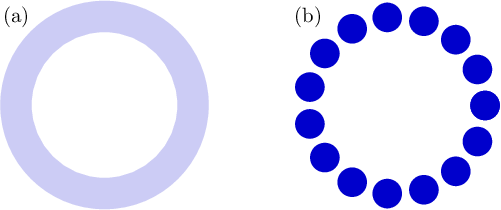}
\caption{%
  Illustration of a rotationally symmetric object with 
  (a) strong rotational interference 
  and (b) vanishing rotational interference.
  If the interference vanishes (b), 
  the quantum system resembles a collection of (rotated) classical objects.
  \label{fig:symm}
}
\end{figure}

The core of this work rests on understanding what is quantum, what is classical, 
and how they are connected. We borrow tools from decoherence theory \cite{JoZe03,Schloss07} and use them for the molecular problem.
Interference is one of the key concepts for understanding quantum behavior. 
If interference vanishes, then the quantum system has  classical-like features. 
In practical terms, 
we have to consider the system's reduced density matrix (RDM).
If its off-diagonal elements (interference terms) are suppressed in some representation,
then the RDM is formally identical with that of a classical mixture (Fig.~\ref{fig:symm}). Based on this mathematical equivalence, we may \emph{say} that a quantum system resembles a classical statistical ensemble.
Science philosophers \cite{FoLoGo16,Ba16} as well as pioneers in decoherence theory \cite{JoZe03} have noticed that suppression of interference between selected states is not a sufficient condition to fully reduce the quantum world to one amenable to a classical treatment. 
It is a necessary one.  In the present work, 
we will follow this common and fruitful interpretation of the quantum-classical relation
that is based on the suppression of the interference terms.

Pursuing decoherence ideas for the molecular problem is not new and 
has always started with the definition of an environment model.
Pfeifer \cite{Pfeifer80,Pf81} considered the vacuum state of 
the electromagnetic field as a general environment 
to model localization of enantiomers.
This model was later debated for several reasons. 
Woolley argued \cite{Wo82} that Pfeifer's non-gauge invariant representation
and linear approximation in the field operators resulted in an artificial localization.
It was also shown that the proposed mechanism fails for non-zero temperature \cite{Wi95}. 
Zeh argued (p. 387 of Ref.~\cite{JoZe03}) that a non-trivial dressing (on its own) does 
not guarantee transformation from a superposition
to an ensemble.

Independently, Claverie and Jona-Lasinio \cite{ClJo86,JoCl86} used external random noise to simulate 
localization in a double-well problem to model molecular chirality. 
Joos and Zeh studied simple scattering models \cite{JoZe85},
and Hornberger and co-workers simulated the stabilization of 
chiral molecules upon collisions \cite{Ho07,BuHo09,TrHo09}.
More recently, change of the environment model parameters were studied on 
the localization of mesoscopic quantum objects \cite{ZhRo16}. 
A systematic and accurate calculation of the decoherence times for 
molecular processes in interaction with a series of environment models
is very interesting and may turn out to be useful for designing better quantum computers (with molecular qubits).

For solving the molecular structure problem, one would like to find the most general possible environment model for a molecule. When we think about the structure of a molecule, we rarely have to specify the corresponding environment.
Most probably, this desire led Pfeifer in 1980 \cite{Pfeifer80} to pick the electromagnetic vacuum state as an environment.

In the present work, instead of setting out to design the most general environment for a molecule, we will take a closer look at the molecular wave function. 
We have said that we did not want to introduce any \emph{a priori} dynamical assumptions in the molecular quantum treatment. But when we have the full molecular wave function at hand, we can, of course, exploit the different properties of the particles during the course of an \emph{a posteriori} analysis of the wave function. 
So, in the present work, we study the spatial coherence or decoherence of the nuclear skeleton induced by the continuous monitoring effect of the electron cloud. This information is encoded in the molecular wave function, and we will write down the formalism that makes this `visible'.

The present study of spatial (de)coherence \emph{within} the molecule adds a missing bit to our earlier understanding \cite{MaHuMuRe11a,MaHuMuRe11b,LuEcLoUg12,BePoLu13,LuIzCoZa16,Sc19} that followed a path proposed by Claverie and Diner \cite{ClDi80}. According to them, elements of the molecular structure can be recognized as nuclear configurations for which the particle density is large.
Large values of the particle density certainly indicate configurations
that are the most probable, but it does not tell us whether an assembly of particles have 
the classical-like features that we know about in chemistry.

In Sec.~\ref{sec:moldec}
we work out the basic decoherence formalism for spatial localization of nuclei in the molecule.
Sec.~\ref{sec:theoshape} makes the formalism specific for the molecular shape 
and Sec.~\ref{sec:numshape} presents numerical results computed from molecular wave functions.

%
%
\clearpage
\section{Molecular decoherence theory \label{sec:moldec}}
\noindent %
When a property of a quantum system is measured, the needle of an idealized measuring
device points to one of the possible outcome values. In an ideal quantum mechanical description of measurement, 
the (macroscopic) experimental setup should be included in a quantum treatment \cite{Neumann55,Primas81}. The quantum states corresponding to the positions of the measuring needle were termed `pointer states' by Zurek \cite{Zurek81} and their apparent classical behaviour is ensured by the decoherence effect of the measuring environment.

Nowadays, the concept of pointer states is used in a more abstract sense without associating an actual experimental setup to them. 
In the pointer basis representation, the off-diagonal elements of the reduced density matrix, which
represent quantum coherence between the states of the system associated to the environment pointer states,
are suppressed.
    
In general, we do not know \emph{a priori} the pointer states of a quantum system in a given environment. For each microscopic environment model, one has to find the proper pointer states that will point to the classical-like (environmentally stabilized) states of the quantum system \cite{BuHo09,Zurek81}. Localization of macroscopic objects in space (\emph{e.g.,} translational localization) is studied in a so-called `direct representation' 
using a set of Dirac delta distributions over the configuration space.

By molecular shape (and internal structure), we understand the localization of the nuclei 
in the three dimensional space. So, our `measuring needles' are functions located in 
the three-dimensional space (`position basis'). Hence, we will be concerned with the `direct representation' of the density matrix. 

\vspace{0.5cm}
Let $|\Psi\rangle$ be a normalized molecular wave function. The associated (pure-state) density operator is denoted by  $\hrho=|\Psi\rangle\langle\Psi|$. 
Let us multiply $\hrho$ both from the left and from the right with the resolution of identity
written in the position basis 
of the electrons and the nuclei, $\chi_r$ and $\chi_R$, respectively.
Hence, the molecular (pure-state) density operator is written as
\begin{align}
  \hat{\rho} 
  &=
  |\Psi\rangle\langle\Psi| 
  \nonumber \\
  &= 
  \hat{I}\cdot |\Psi\rangle\langle\Psi| \cdot\hat{I} 
  \nonumber \\
  &=
  \int \dd r\ \dd R\ 
  |\chi_r \chi_R\rangle \langle \chi_r \chi_R|
  \ \cdot\ %
  |\Psi\rangle\langle\Psi|
  \ \cdot\ %
  \int  \dd r'\ \dd R'\ 
  |\chi_{r'} \chi_{R'}\rangle\langle\chi_{r'} \chi_{R'}| 
  \nonumber\\
%
 &=
  \int \dd r\ \dd R\  \dd r'\ \dd R'\ 
    |\chi_r \chi_R\rangle \langle\chi_{r'} \chi_{R'}|\ 
    \Psi(r,R)
    \Psi^\ast(r',R')
 \label{rdm}
\end{align} 
Since we are interested in the nuclear structure, 
we integrate out the electronic degrees of freedom, and obtain
the nuclear reduced density matrix,
\begin{align}
  \hat{\rho}_\nuc
  &=
  \text{Tr}_\el\ [\hat{\rho} ]
  \nonumber \\
  &=
  \int \dd{r''}\ 
    \langle \chi_{r''}|\Psi \rangle 
    \langle \Psi|\chi_{r''} \rangle 
    \nonumber\\
  &=
  \int \dd{r''}\ 
    \langle \chi_{r''}|
    \left[%
  \int \dd r\ \dd R\  \dd r'\ \dd R'\ 
    |\chi_r \chi_R\rangle \langle\chi_{r'} \chi_{R'}|\ 
    \Psi(r,R)
    \Psi^\ast(r',R')
    \right]
    \chi_{r''} \rangle 
    \nonumber\\    
  %
  &=
  \int \dd R\  \dd R'\ 
  |\chi_R\rangle \langle\chi_{R'}|\   
  \int \dd {r''}\ 
    \Psi({r''},R)
    \Psi^\ast({r''},R') \; ,
  \label{rdm_nuc}
\end{align}
where the element of the reduced density matrix in the position representation is denoted as
\begin{align}
  \rho_\nuc(R,R')
  =
  \langle \chi_R|\hrho_\nuc | \chi_{R'} \rangle 
  = 
  \int \dd {r''}\ \Psi({r''},R)\Psi^*({r''},R') \; .
 \label{rdm_nuc_el}
\end{align}
It is worth making explicit the diagonal and off-diagonal contributions to $\hrho_\nuc$
\begin{align}
  \hrho_\nuc
  =
  &\int \dd R\  
  |\chi_R\rangle \langle\chi_{R}|\   
  \rho_\nuc(R,R)
  \nonumber \\
  &+    
  \int \dd R\  \dd R'\ 
  |\chi_R\rangle \langle\chi_{R'}|\   
  [1-\delta(R-R')]\  
  \rho_\nuc(R,R') \; ,
  \label{eq:rdmnuc_odia}  
\end{align}
where the second term describes the interference between the nuclear structures at $R$ and $R'$.
If the $\rho_\nuc(R,R')$ off-diagonal elements are small, then
\begin{align}
  \hrho_\nuc
  \approx
  &\int \dd R\  
  |\chi_R\rangle \langle\chi_{R}|\   
  \rho_\nuc(R,R)
\end{align}
that is mathematically equivalent with the density matrix of 
a classical mixture of localized structures at $R$ (a mixed state). 
According  to decoherence theory and based on this mathematical equivalence, we may
\emph{say} that the quantum system has classical features corresponding to the $\chi_R$
basis representation that becomes a good pointer basis if $\rho_\nuc(R,R')=0$.
In this case, the $\chi_R$ basis representation is a  pointer basis for the nuclei, and the classical properties associated to a given $\chi_R$ function are dynamically stable. 

We have selected $\chi_R$ as a basis set over the nuclear configuration space (or more precisely, for practical purposes, a dense mesh of Dirac delta distributions over nuclear configurations), because 
we would like to study molecular structure and check whether the interference terms  
are  suppressed between nuclear configurations.
It is necessary to study whether and under which circumstances $\rho_\nuc(R,R')\approx 0$ ($R\neq R'$),
\emph{i.e.,} the nuclear structure will be dynamically stable as a classical property.

\paragraph{Born--Oppenheimer molecular wave function}
It is interesting to note that all arguments are applicable not only for 
the (pre-Born--Oppenheimer) molecular wave function, but also for 
the conventional Born--Oppenheimer (BO) product 
of the electronic and nuclear wave functions, 
$\Psi^{[\text{BO}]}({r},{R})=\Psi_\el({r},{R})\Psi_\nuc({R})$.
In this case, an off-diagonal element of the reduced density matrix is
\begin{align}
  \rho^{\text{[BO]}}_\nuc(R,R')
  &=
  \langle \chi_R |\hat{\rho}_\nuc| \chi_{R'}\rangle \nonumber \\
  &=
    \int \dd{r''}\ 
      \Psi_\el^*({r''},{R'}) \Psi^*_\nuc({R'}) 
      \Psi_\el({r''},{R}) \Psi_\nuc({R})
  \nonumber \\
  &=
  \left[%
    \int \dd{r''}\ 
      \Psi_\el^*({r''},{R'}) 
      \Psi_\el({r''},{R}) 
  \right] 
  \Psi^*_\nuc({R'}) 
  \Psi_\nuc({R}) \; .
  \label{rdm_nuc_BO_el}
\end{align} 
Thus, the interference amplitude between the $R$ and ${R'}$ nuclear structures 
depends on the overlap of the BO electronic wave functions corresponding to the $R$ and $R'$
nuclear configurations. 

\paragraph{All-particle molecular wave function}
The molecular wave function can be represented as linear combination of many-particle electron-nucleus basis functions, most commonly using variants of an explicitly correlated Gaussian 
basis set \cite{rmp13,chemrev13,MaRe12,Ma19review}. Electrons and nuclei are handled in such a  `pre-Born--Oppenheimer' or `all-particle' treatment on an equal footing.
The nuclear reduced density matrix is obtained by direct evaluation of the integral in 
Eq.~(\ref{rdm_nuc_el}). Due to the equivalent treatment of all particles, 
it is straightforward to calculate reduced density matrices corresponding to different kinds of partitioning of the molecule to a subset of $(a,b,c,\ldots)$ particles as the `system' and the remaining $(z,y,x,\ldots)$ particles as the `environment'.
In the present work we use the electron-nucleus partitioning.

\clearpage
\begin{figure}
  \includegraphics[scale=1.]{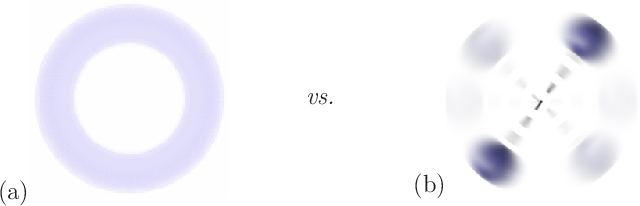}
  \caption{%
    Shell (a) or rotating dumbbell (b)? 
    [E. Mátyus, Mol. Phys. 117, 590 (2019); licensed under a Creative Commons 
    Attribution (CC BY) license.]
    \label{fig:shelldumb}
  }
\end{figure}

\section{Orientational localization: theoretical framework \label{sec:theoshape}}
For a start, let us consider a diatomic molecule in its ground rovibronic state with zero total angular momentum. This is a stationary state and the wave function is spherically symmetric.
The one-particle nuclear density, $\rho_{0,\text{n}}(R,R)$ calculated from this wave function shows that the nuclei (`n') are within a shell around the molecular center of mass, `0' (Fig.~1 of Ref.~\cite{MaHuMuRe11a} and Fig.~2 of Ref.~\cite{MaHuMuRe11b}). 
The two-particle density, $\rho_{0,\text{nn}'}(R,R)$,
is strongly peaked at 180$^\text{o}$ degrees for the included angle of the position vector of the two nuclei, n and n$'$, measured from the center of mass, `0' (Fig.~2 of Ref.~\cite{MaHuMuRe11b}).

The one- and two-particle density functions describe well the internal nuclear structure
but an additional question remains \cite{Ma19review}: 
does the nuclear structure in the ground-state wave function resemble 
a shell with strong quantum coherences among the positions (Fig.~\ref{fig:shelldumb}a)
or 
a rotating dumbbell with a classical-like shape (Fig.~\ref{fig:shelldumb}b)?

We ask this question within the stationary-state, isolated molecule quantum treatment, because we would like to know whether this property is encoded in the molecular wave function itself, or it is induced by the environment. This question can be mathematically studied using the arguments and formalism developed in Sec.~\ref{sec:moldec}. 

\begin{figure}
\includegraphics[scale=1.]{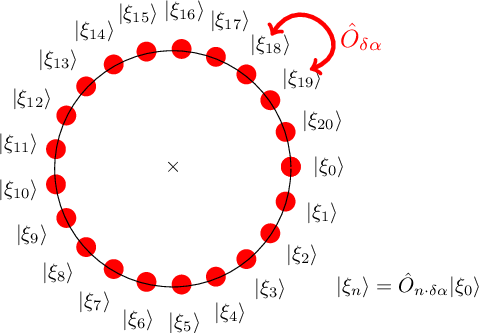}
\caption{%
  Visual representation of the `orientational basis' that we use to measure 
  the emergence of the shape of the nuclear skeleton: 
  Dirac delta distributions are located at molecular structures connected by 
  rotation (shown in 2D).
\label{fig:rot}
}
\end{figure}

So, we have to asses the orientational coherence or decoherence of the nuclei induced by the electrons. For this purpose, we consider nuclear configurations, $R$ and $R'$ that have the same internal structure and are connected by spatial rotation, $R'=\hat{O}_\alpha R$. 

To quantify the interference between rotated structures, 
we will use the $\lbrace \xi_1,\xi_2,\ldots \rbrace$ basis set (Fig.~\ref{fig:rot})
that consists of Dirac delta distributions over a dense mesh 
in the configuration space of rotated nuclear structures,
rather than a general $\chi_R$ configuration basis distributed over
the entire configuration space (Sec.~\ref{sec:moldec}).
Hence, our measuring needles are 
$\xi_\alpha=\chi_{\hat{O}_\alpha R}$ (Fig.~\ref{fig:rot}).
So, in Eq.~(\ref{eq:rdmnuc_odia}) we consider the rotational degrees of freedom for a selected 
$R$ nuclear structure (alternatively, the non-rotational part can be integrated out),
and write the nuclear reduced density matrix operator 
corresponding to the rotated structures as 
\begin{align}
  \hrho^\rot_\nuc
  =
  &\int \dd \alpha\  
  |\xi_\alpha\rangle \langle\xi_{\alpha}|\   
  \rho_\nuc(R_\alpha,R_\alpha)
  \nonumber \\
  &+    
  \int \dd \alpha\  \dd \alpha'\ 
  |\xi_\alpha\rangle \langle\xi_{\alpha'}|\   
  [1-\delta(\alpha-\alpha')]\  
  \rho_\nuc(R_\alpha,R_{\alpha'}) \; ,
\end{align}
where $\dd \alpha$ collects the volume element for three Euler angles and the integrals are understood with the appropriate integration bounds. The 
$R_\alpha=\hat{O}_\alpha R$ is a short notation for the rotated structure.

If the off-diagonal elements of the nuclear reduced density matrix for the rotated structures
($R\neq R_\alpha$), 
\begin{align}
  \rho_\nuc(R,R_\alpha)
  = 
  \int \dd {r''}\ \Psi({r''},R)\Psi^*({r''},R_\alpha) \; ,
 \label{eq:rotoff}
\end{align}
are small with respect to the diagonal elements, 
\begin{align}
  \rho_\nuc(R,R)
  = 
  \int \dd {r''}\ \Psi({r''},R)\Psi^*({r''},R) \; ,
\end{align}
then 
\begin{align}
  \hrho^\rot_\nuc
  \approx
  &\int \dd \alpha\  
  |\xi_\alpha\rangle \langle\xi_{\alpha}|\   
  \rho_\nuc(R_\alpha,R_\alpha) \; ,
\end{align}
\emph{i.e.,} the interference of the rotated nuclear structures is negligible. This 
nuclear reduced density matrix is mathematically identical with that of a classical
mixture of rotated nuclear structures.
This case corresponds to Fig.~\ref{fig:symm}b in a diatomic molecule, the coherence of the rotated structures vanishes, and we may observe the emergence of a dumbbell-like shape of the nuclear skeleton.
If the off-diagonal elements are large with respect to the diagonal ones, then
there is a strong interference among the rotated nuclear structures and we cannot observe any new feature emerging
beyond (within) the quantum mechanical rotational symmetry. In this case, the 
shell-like picture of Fig.~\ref{fig:symm}a is appropriate. 
We note that both cases respect the original rotational symmetry of 
the molecular wave function. Either the classical-like or the quantum-like (as well as all intermediate) cases may be encoded within
the molecular wave function depending on the strength of the interparticle interactions. 
The sole difference is in the suppression or existence of the coherence between pairs of rotated structures within the spherically symmetric molecular wave function.

\vspace{0.5cm}
When do the interference terms get small and the localization of the nuclei 
by the electrons efficient? It happens, if
$\rho_\nuc(R,R_\alpha)$, Eq.~(\ref{eq:rotoff}), is small. 
It is easier to understand the meaning of this condition within 
the Born--Oppenheimer approximation (when it is qualitatively correct), Eq.~(\ref{rdm_nuc_BO_el}):
\begin{align}
  \rho^{\text{[BO]}}_\nuc(R,\hat{O}_\alpha R)
  =
  \left[%
    \int \dd{r''}\ 
    \Psi_\el({r''},{R})
    \Psi_\el^*({r''},{\hat{O}_\alpha R})
  \right] 
  \Psi_\nuc({R})
  \Psi^*_\nuc({\hat{O}_\alpha R})
  \ .
  \label{eq:rhoBO}
\end{align}
If the overlap of the BO electronic wave functions corresponding to the rotated nuclear structures is small, then $\rho^{\text{[BO]}}_\nuc(R,R')$ is also small.

\clearpage
\section{Numerical study of rotational localization of the nuclei in the molecular wave function \label{sec:numshape}}
\noindent We have computed the ground-state wave function for four-particle systems,
including the hydrogen molecule, H$_2$, the positronium molecule, Ps$_2$, the muonium molecule, $\mu_2$, and other H$_2$-like systems, as well as HeH$^+$,
to illustrate the ideas formulated in the preceding sections. 

The molecular wave function was computed using an explicitly correlated Gaussian basis set 
and the QUANTEN computer program \cite{Ma19review}.
In the present work, we use a plain ECG representation similarly to 
Refs.~\cite{MaHuMuRe11a,MaHuMuRe11b} and the center-of-mass-centered translationally invariant
coordinates \cite{MaHuMuRe11b,MaRe12,Ma19review} to evaluate the diagonal and off-diagonal
particle densities referenced to the molecular center of mass.
We have converged the studied structural density features for
the ground state of these systems with zero total angular momentum ($N=0$), natural parity
($p=+1$), and zero spin for the pair of electrons, protons, and positrons ($S_\text{e}=0$, $S_\text{p}=0$). 
The corresponding energies were converged within ca. 1~\% that was sufficient to obtain density plots 
converged within the resolution of the figures.

\begin{figure}
  \includegraphics[scale=1.]{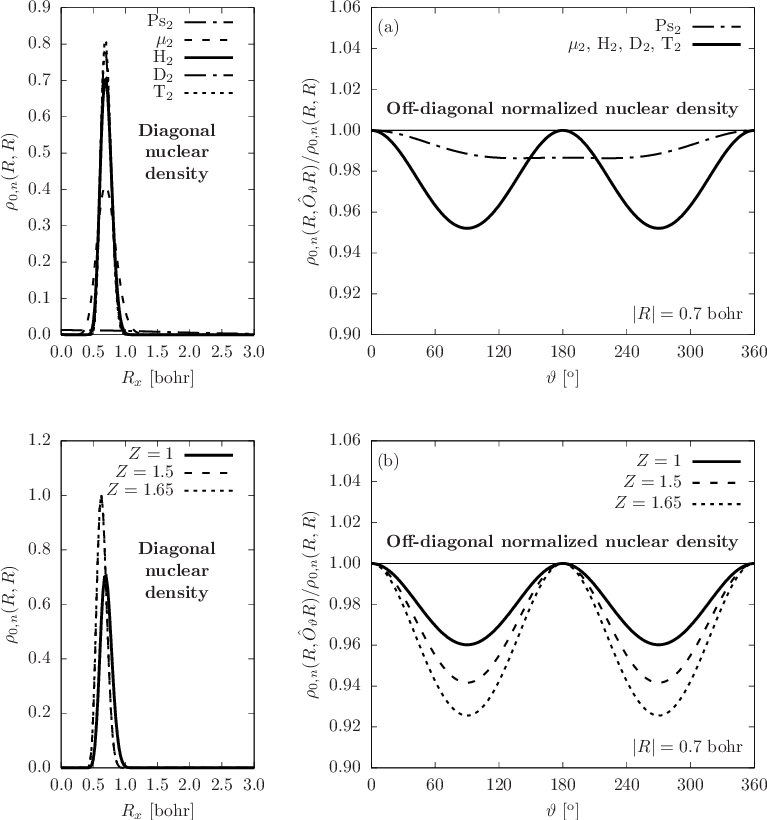}
  \caption{%
    Off-diagonal normalized density matrix elements connecting rotated nuclear structures in H$_2$-type systems. The rotation axis for $\theta$ goes through the molecular center of mass 
    that is the origin of the coordinate system used in the calculations.    
    The diagonal density matrix elements are also plotted.
    (a) Effect of varying the mass of the positively-charged particles (`nuclei'),
    $m_i/m_\text{e}=1, 206.768, 1836.15, 3670.48,$ and $5496.92$
    for Ps$_2$, $\mu_2$, H$_2$, D$_2$, and T$_2$, respectively.
    (b) Effect of varying the electric charge of the proton. \\
    \label{fig:h2ser}
  }
\end{figure}

To answer the question of Ref.~\cite{Ma19review} highlighted in Figure~\ref{fig:shelldumb}, 
we have calculated the off-diagonal elements of the nuclear reduced density matrix
in the rotational basis (Sec.~\ref{sec:theoshape}), 
$\rho_{0,n}(R,\hat{O}_\vartheta R)$, where $\hat{O}_\vartheta$ is the operator for the rotation.
The rotation axis goes through the molecular center of mass and $\vartheta$ parameterizes the rotation angle about this axis.
Figure~\ref{fig:h2ser} shows the effect of the relative mass and electric charge on the orientational (de)coherence 
for a series of H$_2$-type systems
including Ps$_2$, $\mu{_2}$, 
the deuterium and tritium isotopologues of H$_2$ and hypothetical (bound) systems
with an increased nuclear charge.

We may observe in the figures that Ps$_2$ is qualitatively different from the other `homonuclear' systems.
Its diagonal `nuclear' (positron) density, $\rho_{0,n}(R,R)$ has a maximum at the origin ($R=0$~bohr).
Regarding the off-diagonal density, 
it shows a 360$^\text{o}$ periodicity and has a small local maximum at 180$^\text{o}$($\pm i\cdot 360^\text{o}$) and shallow local minima at 139$^\text{o}$ and 221$^\text{o}$
with a 1~\% suppression compared to the maximal (diagonal) value. 

For $\mu_2$, H$_2$, D$_2$, and T$_2$, the diagonal density functions differ but they all have a maximum (of different value for the different systems) near $|R|=0.7$~bohr.
Within the resolution of the figure, the off-diagonal density matrix elements normalized
with the diagonal density value (shown for $|R|$=0.7~bohr, Fig.~\ref{fig:h2ser}a)
cannot be distinguished for the four systems that may be surprising at first sight. 
We may understand this observation by remembering that the suppression is induced by the electrons' measuring effect that depends on 
the electronic structure around the positive particle [Eq. (\ref{eq:rhoBO})] and this is 
very similar in the four systems (the BO approximation is qualitatively correct). 
For this series of systems, the off-diagonal density shows a 180$^\text{o}$ periodicity and it is minimal at 90$^\text{o}$($\pm i\cdot 180^\text{o}$). 
Its minimum value is ca. 5~\% smaller than its maximal (diagonal) value.

Regarding the H$_2$-like systems with an increased nuclear charge (Fig.~\ref{fig:h2ser}b), 
we observe a behavior similar to H$_2$, but a stronger, up to 8~\%, suppression at the 90$^\text{o}$($\pm i\cdot 180^\text{o}$) minima for $Z=1.65$.
This observation is in agreement with the fact that the orientational de-coherence of the nuclei is induced by the monitoring effect of the electrons,
and for an increased nuclear charge this monitoring effect is stronger due to the stronger interaction. (We note that the H$_2$-like with $Z\geq2$ systems are unbound.) 

In summary, we may say that all H$_2$ systems show dominant orientational coherences. Ps$_2$ is an almost perfect quantum ball with 
an only 1\% maximal suppression among the rotated structures. In the H$_2$-like systems, 
the 180$^\text{o}$ periodicity and maximal suppression at 90$^\text{o}$($\pm i\cdot 180^\text{o}$) suggest 
a symmetric dumbbell \emph{shape} (within the spherically symmetric molecular wave function), but
due to the small value of the suppression (5~\%), this dumbbell shape is blurred by quantum coherences.

\begin{figure}
  \includegraphics[scale=1.]{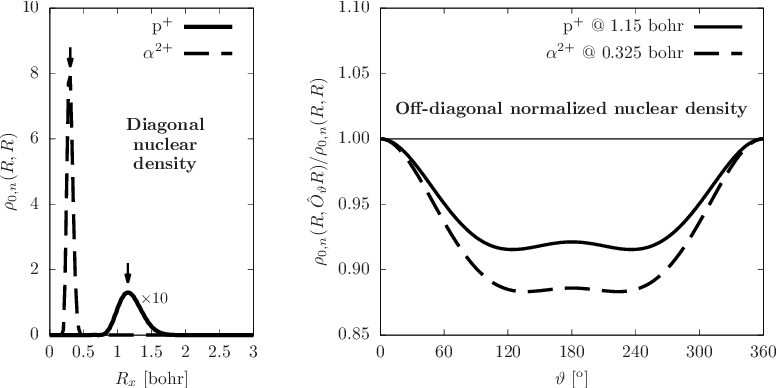}
  \caption{%
    Off-diagonal normalized nuclear density in HeH$^+$ connecting rotated nuclear structures. 
    The diagonal nuclear density is also shown.
    \label{fig:hehp}
    }
\end{figure}

HeH$^+$ is a heteronuclear diatomic molecule with an asymmetric dumbbell shape in the BO theory, while 
its rotational ground-state molecular wave function is, of course, spherically symmetric.
The question arises whether this asymmetry can be recognized within the spherically symmetric molecular wave function. 
Figure~\ref{fig:hehp} shows the diagonal and off-diagonal nuclear density functions. 
The off-diagonal normalized density functions are plotted for $|R|$ values that correspond to the diagonal density maximum value for
the proton and the alpha particles. The off-diagonal density function is periodic by 360$^\text{o}$, 
it has equivalent minima at $\sim 120^\text{o}(\pm i\cdot 360^\text{o}$) and $\sim 240^\text{o}(\pm i\cdot 360^\text{o})$,
and small local maxima at $180^\text{o}(\pm i\cdot 360^\text{o})$ with 
an overall 10~\% suppression for $\vartheta \in[100,300]^\text{o}$ with respect to the maximal (diagonal) density value.

\begin{figure}
  \includegraphics[scale=1.]{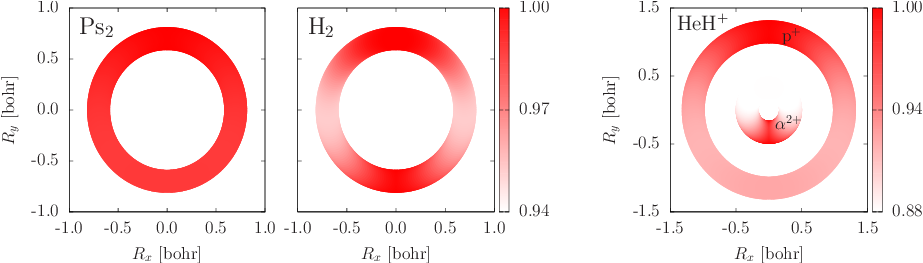}
  \caption{%
    Orientational (de)coherence measured by the suppression of the off-diagonal nuclear density with respect to the diagonal elements,
    $\rho_{0,n}(R,\hat{O}_\vartheta R)/\rho_{0,n}(R,R)$. 
    The figure shows the $xy$ cut of the three-dimensional function (the width of the shells is arbitrarily chosen to visualize the change of the off-diagonal normalized density with the angle).
    The upward (downward) vertical direction correspond to zero rotation angle $\vartheta$
    for the positron and the proton ($\alpha$-particle).
    \label{fig:densplot}
    }
\end{figure}

Figure~\ref{fig:densplot} summarizes the three qualitatively different behavior observed in this work. 
Ps$_2$ is an almost perfect quantum-ball with a maximal 1~\% suppression among rotated structures,
H$_2$ has a symmetric dumbbell shape, while HeH$^+$ has an asymmetric dumbbell shape, 
but both dumbbells are blurred by quantum coherence.

\clearpage
\section{Summary, conclusion, and outlook to future work \label{sec:sum}}
\noindent This work has been devoted to the study of the emergence of 
the classical molecular shape of an isolated molecule 
described by its ground-state rotation-vibration-electronic wave function.
To mathematically formulate this, at first sight, paradoxical problem,
we have studied the orientational decoherence of the nuclear skeleton under
the continuous monitoring effect of the electrons of the molecule, 
serving as the most natural environment for the nuclei. 
Orientational (de)coherence is measured by 
the off-diagonal nuclear reduced density matrix in the (direct) spatial representation.

If the reduced density matrix elements connecting rotated structures are small,
then the nuclear reduced-density matrix is mathematically equivalent to the classical sum 
of rotated nuclear structures, and thus, we may \emph{say} that the nuclei
behave as if they formed a classical-like skeleton with a given shape
that rotates in space.
We have formulated the corresponding equations and computed the 
nuclear reduced density matrix elements over the nuclear configuration space for 
a series of H$_2$-type (homonuclear) systems and for the heteronuclear HeH$^+$.

We observe a small, 10~\% maximal suppression of the off-diagonal density connecting rotated nuclear structures
(due to the measuring effect of the electrons),
and thus we may say that H$_2$ and HeH$^+$ have a symmetric and asymmetric dumbbell-like shape, respectively, that
is blurred by strong quantum coherences.
The decoherence effect in Ps$_2$ is almost negligible and it is seen in this analysis as an almost perfect quantum ball.

We consider molecular rotation and the study of the emergence of the classical molecular
shape (in the restricted sense of suppression of interferences), as a prototypical example for the broader problem of reconstruction of isomers, enantiomers, conformers, rotamers, etc. 
from the molecular wave function (including the electrons and nuclei on an equal footing).
Isomerism, handedness, and related phenomena are linked to polyatomic molecules, 
while already the smallest diatomic molecules have rotational degrees of freedom and 
they are attached with the picture of a classical shape in chemistry.

The present framework uses a spatial basis for the representation of the nuclear
reduced density matrix that can naturally be adapted for studying other space-localized
features. Regarding another challenging aspect of the molecular structure problem, the study of the quantum mechanical indistinguishability vs. the classical distinguishabilty of 
the identical atomic nuclei and implications of the spin-statistics theorem \cite{GrMa18,Bo20} will require perhaps a different framework, but almost certainly a different representation for the reduced density matrix.

The developed ideas will gain more practical significance, if computations without
the Born--Oppenheimer approximation become more widespread, 
which may happen within a new type of quantum chemistry approach perhaps 
on the future hardware of quantum computers \cite{KaJoLoMoAG08}.

Regarding work for the nearer future. We can think about extension of the present work in the following directions.
A next logical step will be to recover the (2-dimensional) triangular shape of
the simplest polyatomic molecule, H$_3^+$ that has an equilateral triangular equilibrium structure
in the Born--Oppenheimer theory.
H$_3^+$ is a system of five spin-1/2 particles (with 2 electrons and 3 protons).
For this reason, the rovibronic ground state of the five-particle Hamiltonian is 
not allowed by the Pauli principle. The lowest energy state that is Pauli allowed has $N=1$ rotational angular momentum, it is the lowest rotationally excited state of the vibrational ground state. 
The lowest Pauli-allowed state with $N=0$ rotational angular momentum is the anti-symmetric stretching fundamental vibration.
So,
the lowest energy $N=1$ state appears to be a good candidate for identifying a near equilateral triangular shape (with small distortion due to rovibrational coupling) in the present framework. 

We have preliminary results for the ground state of H$_2$D$^+$, for which we do not have to deal with complications due to spin statistics, and in its rovibronic ground state ($N=0$), 
we observe a ca. maximal 10~\% suppression among rotated structures. 
We can observe a planar structure in the off-diagonal density plots, but due to the non-equivalent nuclear masses the center of mass (the origin of our computations) is not at the geometrical center of the near
equilateral triangular shape and this complicates the analysis of the results.

For going beyond planar shapes and possibly studying chiral molecules, 
one can rely on the BO approximation (Eq.~(\ref{rdm_nuc_BO_el})), 
since the suppression effect is determined by the electronic structure.
If the BO approximation is qualitatively correct for the system, then the overlap of the rotated BO electronic wave functions should provide a good approximation for the suppression effect that one would observe
for the off-diagonal nuclear reduced density matrix in the molecular wave function (if the computations were feasible).

\vspace{1cm}
\paragraph*{Data Availability Statement}
The data that support findings of this study is included in the paper.

\vspace{1.cm}
\paragraph*{Acknowledgement}
\noindent %
This project was initiated during a Short Term Scientific Mission of the MOLIM COST Action.
EM acknowledges financial support from a PROMYS Grant (no. IZ11Z0\_166525)  
of the Swiss National Science Foundation. 
We thank one of the Reviewers of this article for their insightful comments.


\end{document}